\documentclass[12pt,a4paper]{article}
\usepackage{amsmath}
\usepackage{amssymb}
\usepackage{amsthm}
\usepackage{float}
\usepackage{amsfonts}
\usepackage{graphicx}
\usepackage{verbatim}
\usepackage[left=2cm,right=2cm,top=3cm,bottom=2.5cm]{geometry}
\usepackage[numbers]{natbib}
\usepackage[utf8]{inputenc}
\usepackage[usenames,dvipsnames,svgnames]{xcolor}
\usepackage[colorlinks=true,
      linkcolor=red,
      urlcolor=gray,
      citecolor=blue]{hyperref}
\newcommand{\be}{\begin{equation}}
\newcommand{\ee}{\end{equation}}
\newcommand{\ber}{\begin{eqnarray}}
\newcommand{\eer}{\end{eqnarray}}
\newcommand{\bern}{\begin{eqnarray*}}
\newcommand{\eern}{\end{eqnarray*}}
\newcommand{\beast}{\begin{equation*}}
\newcommand{\eeast}{\end{equation*}}

\usepackage{float}
\def\case#1/#2{\frac{#1}{#2}}

\def \D {\tilde{\nabla}}
\def\la {\langle}
\def\ra {\rangle}
\newcommand{\sfrac}[2]{{\textstyle{#1\over#2}}}
\def \ep {\varepsilon}
\def\tl{\tilde}
\def\rd {\displaystyle{\cdot}}
\def\ts {\textstyle}
\def\myalign#1{%
  \def\trule{\noalign{\smallskip\hrule\medskip}}
  \def\nebc{\nearrow\bigcup}
  \def\sebc{\searrow\bigcup}
  \def\pminf{{}_{-\infty}|^{+\infty}}
  \let\Inf\infty
  \def\amp{&} 
  \vbox{\mathsurround0pt\openup1\jot
    \halign{%
      &$\displaystyle##\hfil\tabskip0pt$&\amp##\tabskip1em\crcr
      \noalign{\hrule height1pt\smallskip}#1\noalign{\smallskip\hrule height1pt}\crcr}}}
      
\begin{document}
\begin{center}
\textbf{On Chaplygin models in $f(G)$ gravity}
\end{center}
\hfill\\
Fidele Twagirayezu$^{1}$, Abraham Ayirwanda$^{1}$, Albert Munyeshyaka$^{2}$, Solange Mukeshimana$^{3}$, Joseph Ntahompagaze$^{1}$, and Leon Fidele Ruganzu Uwimbabazi$^{3}$\\
\hfill\\
$^{1}$Department of Physics, College of Science and Technology, University of Rwanda, Rwanda\;\;\; \; \;
\hfill\\
$^{2}$Department of Physics, Mbarara University of Science and Technology, Mbarara, Uganda
\;\;\; \; \;\hfill\\
$^{3}$Department of Mathematics, College of Science and Technology, University of Rwanda, Rwanda.
\;\;\; \; \;\;\; \; \hfill\\
\hfill\\
Correspondence: ntahompagazej@gmail.com \;\;\;\;\;\;\;\;\;\;\;\;\;\;\;\;\;\;\;\;\;\;\;\;\;\;\;\;\;\;\;\;\;\;\;\;\;\;\;\;\;\;\;\;\;\;\;\;\;\;\;\;\;\;\;\;\;\;\;\;\;\;\;\;\;\;\;\;\;\;\;\;\;\;\;\;\;\;\;\;\;\;\;\;\;\;\;\;\;\;\;\;\;\;\;\;\;\;\;
\begin{center}
\textbf{Abstract}
\end{center}
The current work treats cosmological perturbation in a mixture of standard matter, Chaplygin gas as well as Gauss-bonnet fluids using a 1+3 covariant approach in the context of modified $f(G)$ gravity. We define the gradient variables to obtain linear perturbation equations.  After scalar and redshift transformations, we consider both an original Chaplygin and generalized Chaplygin gas models  under Gauss-bonnet gravity. For pedagogical purposes, the consideration of polynomial $f(G)$ gravity model was used to  solve the perturbation equations for short- and long- wavelength modes and investigate the late time evolution. The numerical solutions were obtained. The results show that the energy overdensity perturbations decay with an increase in redshift. The treatment recovers GR results under limiting cases.\\
\hfill\\
\textit{keywords:} Cosmology; Chaplygin gas; Covariant perturbations; Dark energy

 \section{Introduction}
Recent studies in modern cosmology have revealed information that the present universe is undergoing a cosmic acceleration \cite{bengaly2020evidence,riess1998observational,perlmutter1999measurements}. Such an acceleration according to standard model of cosmology is caused by a fluid with a negative pressure \cite{copeland2006dynamics,bamba2012dark,wang2014post}. This model is built on the Einstein's theory of relativity known as General Relativity (GR), which was so far a fundamental and a successful theory of gravity for several years \cite{vishwakarma2016einstein}. However, GR suffers to explain the recent observed cosmic acceleration at least without invoking the cosmological constant \cite{borges2008evolution,tegmark2004cosmological,kumar2021remedy}. Recently, it was shown that one can consider Chaplygin gas model, a model with a negative pressure by construction, to employ it under cosmological set up. This can help to study and possibly explain the late cosmic acceleration as well as other cosmological features raised through modern observational cosmology \cite{hough2021confronting}. Several works have treated the Chaplygin gas model in GR \cite{saadat2013viscous,dev2003cosmological,debnath2004role}, in $f(R)$ gravity \cite{elmardi2016chaplygin,sami2017inflationary}, $R$ being Ricci scalar, in $f(T)$ gravity \cite{sahlu2019chaplygin}, $T$ being the torsion scalar, to name a few. It was shown that the Chaplygin gas models can contribute to the understanding of the evolution of the universe through perturbation analysis \cite{hough2021confronting}. There are different types of Chaplygin gas models. For instance, the original and generalized Chaplygin gas model was treated in \cite{gorini2005chaplygin} where the scalar field contribution were discussed. Other exploration of the same model was previously done in \cite{nojiri2005inhomogeneous,elmardi2016chaplygin,sami2017inflationary,gorini2003can}.\\
\hfill\\
In the present work, a mixture of a Chaplygin gas fluid, standard matter fluid as well as Gauss-bonnet fluid, as non-interacting fluids, is considered using a $1+3$ covariant approach. The motivation behind the choice of $1+3$ covariant approach is that the perturbation equations that describe the dynamical evolution of the universe and predicting the large scale structures formation in late time are easily found using that approach rather than metric perturbation. In addition to that, the advantage to consider the $1 + 3$ covariant formalism over the metric formalism relies in the fact that it does not leave physical modes in the evolution of perturbations \cite{ellis1989covariant,dunsby1992cosmological}. The 1+ 3 covariant approach being the way of dividing the space-time into foliated hyper-surfaces and a perpendicular 4-vector-field, it was introduced by Ellis and Bruni in 1989 \cite{ellis1999cosmological} focusing on GR. Progressively after the recent discovery of cosmic acceleration, it gained much attention to extent that it was employed to different modified gravity theories \cite{abebe2012covariant,abebe2013large}. \\
 \hfill\\
The next part of this paper is organized as follows: in Section \ref{section2}, the presentation of the mathematical framework
where one describes the $1+3$ approach, defines the vector gradients and linear evolution of the variables, is done. The section \ref{section3} covers the  linear evolutions equations, and presents the scalar perturbation equations in redshift space. In Section  \ref{section4}, we presents  the asymptotic description of the system with the consideration of both  long-and short- wavelength modes in both dust and radiation epochs respectively and present the numerical results of the perturbation equations by considering  both $GR$ limit and the system of matter chaplygin gas and Gauss-Bonnet fluids. The section \ref{section5} is devoted to discussions and conclusions. The adopted spacetime signature is $(- + + +)$ and unless stated otherwise, we have used the convention $8\pi G = c = 1$, where $G$ is the gravitational constant and $c$ is the speed of light.
\section{Mathematical Framework}\label{section2}
 In this section, we  present the mathematical aspects to describe the cosmic evolution. In this regard, we first of all define the vector gradient variables of individual fluid and find how it evolves. The action for modified Gauss-Bonnet gravity is given as \cite{nojiri2005modified,cognola2006dark,li2007cosmology,nojiri2011unified,garcia2011energy,nojiri2017modified}
\begin{eqnarray}
 S= \frac{1}{2} \int d^{4}x\sqrt{-g}\Big[R+f(G)+2\mathcal{L}_m\Big],
\end{eqnarray}
 with Kappa ($\kappa$)  is assumed to equals to $1$ and $R$ is the Ricci scalar and $G$ is the Gasuu-Bonnet parameter.  The modified Einstein equation  becomes
  \begin{eqnarray}
   && R_{\mu\nu}-\frac{1}{2}g^{\mu\nu}R= T^{m}_{\mu\nu}+\frac{1}{2}g^{\mu\nu}f-2f'RR^{\mu\nu}+4f'R^{\mu}_{\lambda}R^{\nu\lambda}-2f'R^{\mu\nu\sigma\tau}R^{\lambda\sigma\tau}_{b}\nonumber\\&&-4f'R^{\mu\lambda\sigma\nu}R_{\lambda\sigma} +2R\bigtriangledown^{\mu}\bigtriangledown_{\nu}f'-2Rg^{\mu\nu}\bigtriangledown^{2}f'-4R^{\nu\lambda}\bigtriangledown_{\lambda}\bigtriangledown^{\mu}f'\nonumber\\&&-4R^{\mu\lambda}\bigtriangledown_{\lambda}\bigtriangledown^{\nu}f'+4R^{\mu\nu}\bigtriangledown^{2}f' +4g^{\mu\nu}R^{\lambda\sigma}\bigtriangledown_{\lambda}\bigtriangledown_{\sigma}f'-4R^{\mu\lambda\nu\sigma}\bigtriangledown_{\lambda}\bigtriangledown_{\sigma}f',
   \label{eq2}
  \end{eqnarray}
where $f\equiv f(G)$ and $f'=\frac{\partial f}{\partial G}$ and $T^{m}_{\mu\nu}$ is the energy momentum tensor of the matter fluid  (photons, baryons, cold dark matter, and light neutrinos) with  $G$ is  given as $R^{2}-4R_{\mu \nu}R^{\mu \nu}+R_{\mu\nu\alpha \beta} R^{\mu\nu\alpha \beta}$, $R^{\mu \nu}$ is the Ricci tensor and $R^{\mu\nu\alpha \beta}$ is the Riemann tensor.
The energy-momentum tensor of matter fluid forms is given by
$ T_{\mu \nu} = \rho u_{\mu}u_{\nu} + p h_{\mu\nu}+ q_{\mu}u_{\nu}+ q_{\nu}u_{\mu}+\pi_{\mu \nu}$
where $\rho$, $p$, $q_{\mu}$ and $\pi_{\mu\nu}$ are the energy density, isotropic
pressure, heat flux and anisotropic pressure of the fluid respectively.
The quantities $\pi_{\mu\nu}$, $q^{\mu}$, $\rho$ and $p$ are reffered  to as dynamical quantities.
The quantities $\sigma_{\mu\nu}$, $\varpi_{\mu\nu}$, $\theta$ and $A_{\mu}$ are reffered as kinematical quantities.
The consideration of  standard matter fluids (dust, radiation, etc), Chaplygin gas fluid and Gauss-Bonnet contributions leads  us to define  the total energy density, isotropic  pressures,  as
\begin{equation}
 \rho_{t}=\rho_{m}+\rho_{ch}+\rho_{G},~ p_{t}=p_{m}+p_{ch}+p_{G},
 \label{eq3}
\end{equation}
where \cite{hough2021confronting,li2007cosmology,garcia2011energy,kamenshchik2001alternative,munyeshyaka2021cosmological,venikoudis2022late}
\begin{eqnarray}
 && \rho_{G}=\frac{1}{2}(f'G-f)-24f''\dot{G},
 \label{eq4}\\
 && p_{G}=\frac{1}{2}(f-f'G)+\frac{G\dot{G}}{3H}f''+4H^{2}\ddot{G}f''+4H^{2}\dot{G}^{2}f''',\\
 && \rho_{ch}=\Big[A+\frac{C}{a^{3(1+\alpha)}}\Big]^{\frac{1}{1+\alpha}},\\
 && p_{ch}=-\frac{A}{(\rho_{ch})^{\alpha}}.
 \label{eq7}
\end{eqnarray}
We assume a spatially flat Friedman-Robert-Walker(FRD)  universe,
\begin{equation}
 ds^{2}=-dt^{2}+a^{2}\left(dx^{2}+dy^{2}+dz^{2}\right),
\end{equation}
  so that  the equation corresponding to the Friedmann equation is modified as
\begin{eqnarray}
 &&3H^{2}=\frac{1}{2}\left( Gf'-f-24\dot{G}H^{3}f''\right)+\Big[A+\frac{C}{a^{3(1+\alpha)}}\Big]^{\frac{1}{1+\alpha}}+\rho_{m}\;,
 \label{eq9} \\
 &&  G=24H^{2}(\dot{H}+H^{2})\;,
 \label{eq10}\\
  &&  R=6(\dot{H}+2H^{2})\;,
  \label{eq11}
\end{eqnarray}
where $A$, $C$ are arbitrary constants, $H=\frac{\dot{a}}{a}$ is the Hubble parameter and $a$ is the scale factor.
The  energy density and pressure in the modified Gauss-Bonnet gravity are presented as
\begin{eqnarray}
&& \rho_{t}=3H^{2}\;,
 \label{eq12}\\
&& p_{t}=-(3H^{2}+2\dot{H})\;.
 \label{eq13}
\end{eqnarray}
\subsection{The 1+3 Covariant approach in the context of $f(G)$ gravity}
The $1 + 3$ covariant decomposition is a framework used in describing the linear evolution of the cosmological perturbations \cite{sahlu2020scalar,munyeshyaka20231+}.
In this approach, a fundamental observer divides space-time into hyper-surfaces and a perpendicular 4-velocity field vector where $1 + 3$ indicates the number of dimensions involved in each slice \cite{clarkson2007covariant}.That is to mean that manifold geometry of the GR is discribed in four dimensinal spacs (ie,. time and space).  One of the importance of the $1 + 3$ covariant approach is to identity a set of covariant variables which describe the inhomogeneity and anisotropy of the universe \cite{ntahompagaze2018study}. In this context, we define a four-vector coordinateas function of cosmological time ($x^{\mu}=x^{\mu}(\tau)$) that  labels the comoving distance along a world-line and the corresponding velocitiy given as :
\begin{equation}
 u^{\mu}= \frac{dx^\mu}{d\tau}
\end{equation}
The projection tensor, $h_{\alpha\beta}$ into the three dimensional and orthogonal to $u^{\mu}$, satisfy the following condition:
\begin{eqnarray}
 && h_{\alpha\beta}=g_{\alpha\beta}+u_{\alpha}u_{\beta}\Rightarrow h^{\alpha}_{\beta}h^{\beta}_{\gamma}=h^{\alpha}_{\gamma},\\
 &&h^{\alpha}_{\alpha}= 3, h_{\alpha\beta}u^{\beta}=0.
\end{eqnarray}
  The covariant derivative of the four-velocitiy in terms of its kinematic quantities \cite{ntahompagaze2020multifluid} is given by:
\begin{eqnarray}
 \tilde{\bigtriangledown}_{a}u_{b}= \frac{1}{3}h_{ab}\tilde{\theta} +\tilde{\sigma}_{ab}+\tilde{\omega}_{ab}-u_{a}{\dot{\tilde{u}}}_{b}
\end{eqnarray}
Where $\theta$, $\tilde{\sigma}_{ab}$, $\tilde{\omega}_{ab}$, $\dot{\tilde{u}}_{b}$, are: the volume expansion,sheartensor, vorocity tensor and four-acceleration respectively. The Hubble parameter is related to $\theta$ as $\theta=3H$. Assume the fluids in our  consideration are irrotational (ie.,$\tilde{\omega}_{ab} =0$) and shear-free (i.e $\tilde{\sigma}_{ab}=0$), the rate of expansion is given by the Raychaudhuli and conservation  equations as:
\begin{eqnarray}
 &&\dot{\theta}= -\frac{\theta^{2}}{3}-\frac{1}{2}(\rho_{t}+p_{t})+\tilde{\bigtriangledown}^{a}\dot{u}_{a},\\
&& \dot{\rho}= -\theta(\rho+p),\\
 &&\tilde{\bigtriangledown}_{a}p-(\rho_{t}+p_{t})\dot{u}_{a}=0.
\end{eqnarray}
\subsection{General fluids description}
In this part, we assume a non-interacting  matter fluids with both chaplygin gas and Gauss-Bonnet fluids in the entire Universe where the growth of the energy overdensity fluctuations contribute to the large scale structure formation.
We consider an homogenous and expanding (FRW) cosmological background where it is possible to define the spatial gradient  of the Gauge invariant variables  as $
D^{m}_{a}=\frac{a \tilde{\nabla}^{a}\rho_{m}}{\rho_{m}},
 Z_{a}=a\tilde{\nabla}^{a}\theta,
 D_{a}^{ch}=\frac{a \tilde{\nabla}_{a}\rho_{ch}}{\rho_{ch}}\;,\\
 {D}^{G}_{a}= \frac{a\tilde{\nabla_{a}}\rho_{G}}{\rho_{G}},
\mathcal{G}_{a}=a\tilde{\nabla}_{a}{G}, \mathsf{G}_{a}=a\tilde{\nabla}_{a}\dot{G}\;$.
The subscipts $m$, $G$ and  $ch$  stand for matter, Gauss-Bonnet fluid, Chaplygin gas fluid contributions respectively.
  All the defined gradient variables will be considered to develop a system of cosmological perturbation equations in the context of chaplygin-gas assisted $f(G)$ gravity.
 \section{Linear evolution equations}\label{section3}
Taking the first derivative with respect to time of the defined gradient variables, we get:
\begin{eqnarray}
&&\dot{D}_{m}^{a}=-\left(1+w_{m}+\frac{(1+w_{m})}{(1+w_{t})\rho_{t}}\frac{G\dot{G}}{\theta^{2}}\right)Z{a}+\frac{(1+w_{m})}{(1+w_{t})}\frac{\rho_{m}}{\rho_{t}}\omega_{m}\theta D_{m}^{a}+\frac{(1+w_{m})}{(1+w_{t})\rho_{t}}w_{chp}\rho_{chp}D_{chp}^{a} \nonumber\\
&&+\frac{(1+w_{m})}{(1+w_{t})\rho_{t}}\rho_{chp}a\nabla^{a}w_{chp}
+\frac{(1+w_{m})}{(1+w_{t})\rho_{t}}(\frac{1}{2}(1-f'-Gf'')+G\dot{G}f'''+\frac{\dot{G}f''}{G^{2}})\mathcal{G}_{a}\nonumber\\
&&+\frac{(1+w_{m})}{(1+w_{t})\rho_{t}}\frac{Gf''}{\theta}G_{a}
\label{eq21}\\
&& \dot{\Delta_{ch}}= -[1+\frac{G\dot{G}}{\theta(1+w_{t})\rho_{t}}](1+w_{ch})Z+\frac{\theta (1+w_{ch})}{(1+w_{t})\rho_{t}}w_{m}\rho_{m}\Delta_{m}+(1+\rho_{ch})\frac{\theta (1+w_{ch})}{(1+w_{t})\rho_{t}}w_{ch}\rho_{ch}\Delta_{ch}\nonumber\\&&-a\theta\tilde{\nabla}_{a}w_{ch}+\frac{\theta (1+w_{ch})}{(1+w_{t})\rho_{t}}[\frac{1}{2}(1-f'-Gf'')+G\dot{G}f''+\frac{\dot{G}f'''}{\theta^{2}}]\mathsf{G}_{a}+\frac{\theta (1+w_{ch})}{(1+w_{t})\rho_{t}}Gf''	\mathsf{G},\\
 \label{eq22}
 &&\dot{\mathsf{G}}_{a}=(\frac{\dddot{G}}{\dot{G}}-\frac{\ddot{G}{G}f''}{(1+w_{t})\rho_{t}\theta})\mathsf{G}_{a}-\frac{\ddot{G}\dot{G}{G}}{(1+w_{t})\rho_{t}\theta^{2}}Z_{a}-\frac{\ddot{G}}{(1+w_{t})\rho_{t}}w_{m} \rho_{m} D_{a}^{m}-\frac{\ddot{G} \rho_{ch}w_{ch}D_{a}^{ch}}{(1+w_{t})\rho_{t}}\nonumber\\&&-\frac{\ddot{G}\rho_{ch} a\nabla_{a}w_{ch}}{(1+w_{t})\rho_{t}}-\frac{\ddot{G}}{2(1+w_{t})\rho_{t}}(1-f'-Gf'')\mathcal{G}_{a}-\frac{\ddot{G}\dot{G}}{(1+w_{t})\rho_{t}}(Gf''+\frac{f''}{\theta^{2}})\mathcal{G}_{a}
  \label{eq23}
   \end{eqnarray}
\begin{eqnarray}
&&\dot{\mathcal{G}}_{a}= (1-\frac{\dot{G} G f''}{(1+w_{t})\rho_{t} \theta})\mathsf{G}_{a} +\frac{\dot{G}^{2} G}{(1+w_{t})\rho_{t} \theta^{2}} Z_{a} -\frac{\dot{G} w_{m} \rho_{m} D_{a}^{m}}{(1+w_{t})\rho_{t}} -\frac{\dot{G} \rho_{ch}w_{ch}D_{a}^{ch}}{(1+w_{t})\rho_{t}}- \frac{\dot{G}\rho_{ch} a\nabla_{a}w_{ch}}{(1+w_{t})\rho_{t}}+\nonumber\\&&\Big[\frac{\dot{G}}{2(1+w_{t})\rho_{t}}(f'+Gf''-1)-\frac{\dot{G}^{2}}{2(1+w_{t})\rho_{t}}(Gf''+\frac{f''}{\theta^{2}})\Big]\mathcal{G}_{a},
\label{eq24}
\end{eqnarray}
\begin{eqnarray}
&&\dot{Z}_{a}=-\Big[\frac{1}{2}(1+3w_{m})\rho_{m}+\big(\frac{1}{3}\theta^{2}+\frac{1}{2}(1+3w_{t})\rho_{t}\big)\frac{w_{m}\rho_{m}}{(1+w_{t})\rho_{t}}\Big]D^{m}_{a}\nonumber\\
&&-\Big[\frac{1}{2}(1+3w_{chp})\rho_{chp}-\big(\frac{1}{3}\theta^{2}+\frac{1}{2}(1+3w_{t})\rho_{t}\big)\frac{w_{chp}\rho_{chp}}{(1+w_{t})\rho_{t}}\Big]D^{chp}_{a}\nonumber\\
&&-\Big[\frac{3}{2}\rho_{chp}-\big(\frac{1}{3}\theta^{2}+\frac{1}{2}(1+3w_{t})\rho_{t}\big)\frac{\rho_{chp}}{(1+w_{t})\rho_{t}}\Big]a\tilde{\bigtriangledown}_{a}w_{chp}\nonumber\\
&&+\Big[1-3f''-12H^{3}\dot{G}f^{iv}+12H\Big(H(\ddot{G}f'''+\frac{\dddot{G}}{G}f''+\dot{G}^{2}f^{iv})+2H^{3}f''\Big)\nonumber\\
&&+\big[\frac{1}{3}\theta^{2}+\frac{1}{2}(1+3w_{t})\rho_{t}\big]\frac{1}{(1+w_{t})\rho_{t}}\Big(\frac{1}{2}(1-f'-Gf'-Gf'')+G\dot{G}f'''+\frac{\dot{G}f''}{G^{2}}\Big)\Big]\mathcal{G}_{a}\nonumber\\
&&+\Big(24\dot{G}f'''-12H^{3}f'''+\big(\frac{1}{3}\theta^{2}+\frac{1}{2}(1+3w_{t})\rho_{t}\big)\frac{Gf''}{\theta(1+w_{t})\rho_{t}}\Big)\mathbf{G}_{a}\nonumber\\
&&+\Big[8H\left(f''\ddot{G}+f'''\dot{G}^{2}+3Hf'-\frac{3}{2}f'''H\dot{G}\right)-\frac{2}{3}\theta-\big(\frac{1}{3}\theta^{2}+\frac{1}{2}(1+3w_{t})\rho_{t}\big)\frac{G\dot{G}}{\theta^{2}(1+w_{t})\rho_{t}}\Big]Z_{a}\nonumber\\&& -\frac{1}{(1+w_{t})\rho_{t}}\bigtriangledown^{2}\Big[w_{m}\rho_{m}D^{m}_{a}+w_{chp}\rho_{chp}D^{chp}_{a}+\rho_{chp}a\tilde{\bigtriangledown}^{a}w_{chp}\Big]\nonumber\\
&&-\frac{1}{(1+w_{t})\rho_{t}}\bigtriangledown^{2}\Big[(\frac{1}{2}(1-f'-Gf'')+G\dot{G}f'''+\frac{\dot{G}f''}{G^{2}})\mathcal{G}_{a}+\frac{Gf''}{\theta}G_{a}-\frac{G\dot{G}}{\theta^{2}}Z_{a}\Big],
\label{eq25}\\
&&\dot{D}^{G}_{a}=-\Big[\frac{4H\theta}{\rho_{G}}\Big(2H\dot{G}f''' +(2\dot{H}-2H^{2})f''\Big)-\theta\Big(4H^{2}\ddot{f}'+4H(2\dot{H}\nonumber\\&&-H^{2})\dot{f}'\Big)\Big(\frac{Gf''}{\theta(1+w_{t})\rho_{t}\rho_{G}}\Big)\Big]\mathsf{G}_{a}
+4H\Big[H(\ddot{G}f'''+\frac{f''\dddot{G}}{\dot{G}}+\dot{G}^{2}f''')+(2\dot{H}-H^{2})f''\nonumber\\&&+\theta\Big(4H^{2}\ddot{f}'+4H(2\dot{H}-H^{2})\dot{f}'\Big)\Big(\frac{\frac{1}{2}+G\dot{G}f''+\frac{\dot{G}f'''}{\theta^{2}}}{(1+w_{t})\rho_{t}\rho_{G}}\Big)\Big]\mathcal{G}_{a} +\frac{1}{3\rho_{G}}\Big[\Big(2H(f''\ddot{G}+f'''\dot{G}^{2})\nonumber\\
&&-8Hf''\dot{G}(\frac{G}{12H^{3}}+3H)+(2\dot{H}-H^{2})f''\dot{G}-3\Big(4H^{2}\ddot{f}'+4H(2\dot{H}-H^{2})\dot{f}'\Big)\nonumber\\
&&+\theta(4H^{2}\ddot{f}'-4H(2\dot{H}-H^{2})\dot{f}')(\frac{G\dot{G}\theta^{2}}{(1+w_{t})\rho_{t}\rho_{G}})\Big)\nonumber\\ &&+3\theta(4H^{2}\ddot{f}'+4H(2\dot{H}-H^{2})\dot{f}')\Big(\frac{(1-f'-Gf'')G\dot{G}}{\theta^{2}(1+w_{t})\rho_{t}}\Big)\Big]Z_{a}
\nonumber\\
&&+\frac{\theta}{{\rho_{G}}}\Big(4H^{2}\ddot{f}'+4H(2\dot{H}-H^{2})\dot{f}'\Big) D^{G}_{a}\nonumber\\
&&+\theta\Big(-4H^{2}\ddot{f}'+4H(2\dot{H}-H^{2})\dot{f}'\Big)\frac{w_{m}\rho_{m}(1-f'-Gf'')}{(1+w_{t}\rho_{t}\rho_{G})}D^{m}_{a}\nonumber\\
&&+\theta\Big(4H^{2}\ddot{f}'+4H(2\dot{H}-H^{2})\dot{f}'\Big)\Big(\frac{w_{ch}\rho_{ch}}{(1+w_{t})\rho_{t}\rho_{a}}\Big)a\tilde{\nabla}_{a}w_{ch}
\label{eq26}
 \end{eqnarray}
 We have used
 \begin{equation}
\dot{u}_{a}=-\frac{\tilde{\nabla}^{a}p_{t}}{\rho_{t}+p_{t}}
 \end{equation}
  \begin{equation}
 a\dot{u}_{a}=-\frac{1}{(1+w_{t})\rho_{t}}[w_{m}\rho_{m}D^{a}_{m}+w_{chp}\rho_{chp}D^{a}_{chp}+a\rho_{chp}\tilde{\nabla}^{a}w_{chp}+\tilde{\nabla}^{a}p_{G}]\;,
\end{equation}
These linear vector evolution equations (eq. \ref{eq21}-\ref{eq26})  can be represented in redshift space  after applying scalar and harmonic decomposition methods. It can be shown that the scalar perturbation equations  for the matter, Chaplygin gas and Gauss-Bonnet energy densities in redshift space evolve as
  \cite{munyeshyaka2022perturbations,sahlu2020perturbations,venikoudis2022late,munyeshyaka20231+}
  \begin{equation}
 \begin{split}
 &\Delta'_{m}=\frac{1}{(1+z)H}\left(1+w_{m}+\frac{(1+w_{m})}{(1+w_{t})\rho_{t}}\frac{G\big(-(1+z)H)\big)G'}{\theta^{2}}\right)Z\nonumber\\
 &-\frac{1}{(1+z)H}\frac{(1+w_{m})}{(1+w_{t})}\frac{\rho_{m}}{\rho_{t}}\omega_{m}\theta \Delta_{m}
 +\frac{1}{(1+z)H}\alpha \frac{(1+w_{m})}{(1+w_{t})\rho_{t}}w_{chp}\rho_{chp}\Delta_{chp}\nonumber\\
 &+\frac{(1+w_{m})}{(1+w_{t})\rho_{t}}\frac{Gf''}{\theta}\mathsf{G}
-\frac{1}{(1+z)H}\frac{(1+w_{m})}{(1+w_{t})\rho_{t}}\Big(\frac{1}{2}(1-f'-Gf'')G(1+z)HG'f'''\nonumber\\
&-\frac{(1+z)HG'f''}{G^{2}}\Big)\mathcal{G}
 \label{eq29}
 \end{split}
\end{equation}
\begin{eqnarray}
&& Z'=-\frac{1}{(1+z)H}\Big[-\frac{1}{2}(1+3w_{m})\rho_{m}-\big(\frac{1}{3}\theta^{2}+\frac{1}{2}(1+3w_{t})\rho_{t}\big)\frac{w_{m}\rho_{m}}{(1+w_{t})\rho_{t}}-\frac{w_{m}\rho_{m}}{(1+w_{t})\rho_{t}}\frac{k^{2}}{a^{2}}\Big]\Delta_{m}+\nonumber\\
&&-\frac{1}{(1+z)H}\Big[-\frac{1}{2}(1+3w_{chp})\rho_{chp}+\big(\frac{1}{3}\theta^{2}+\frac{1}{2}(1+3w_{t})\rho_{t}\big)\frac{w_{chp}\rho_{chp}}{(1+w_{t})\rho_{t}}\nonumber\\
&&\Big[\frac{3}{2}\rho_{chp}-\big(\frac{1}{3}\theta^{2}+\frac{1}{2}(1+3w_{t})\rho_{t}\big)\frac{\rho_{chp}}{(1+w_{t})\rho_{t}}\Big](\alpha+1)w_{chp}-\frac{\big(w_{chp}\rho_{chp}-(\alpha+1)w_{chp}\big)}{(1+w_{t})\rho_{t}}\frac{k^{2}}{a^{2}}\Big]\Delta_{chp}\nonumber\\&&-\frac{1}{(1+z)H}\Big[1-3f''-12H^{3}b_{1}f^{iv}+12H\Big(H(b_{2}f'''+\frac{b_{3}}{b_{1}}f''+b_{1}^{2}f^{iv})+2H^{3}f''\Big)\nonumber\\
&&+\big(\frac{1}{3}\theta^{2}+\frac{1}{2}(1+3w_{t})\rho_{t}\big)\frac{1}{(1+w_{t})\rho_{t}}\Big(\frac{1}{2}(1-f'-Gf'-Gf'')+Gb_{1}f'''+\frac{b_{1}f''}{G^{2}}\Big)\nonumber\\&& -\frac{1}{(1+w_{t})\rho_{t}}\frac{k^{2}}{a^{2}}\Big(\frac{1}{2}(1-f'-Gf'')+Gb_{1}f'''+\frac{b_{1}f''}{G^{2}}\Big)\Big]\mathcal{G}\nonumber\\
&&-\frac{1}{(1+z)H}\Big(24b_{1}f'''-12H^{3}f'''+\big(\frac{1}{3}\theta^{2}+\frac{1}{2}(1+3w_{t})\rho_{t}\big)\frac{Gf''}{\theta(1+w_{t})\rho_{t}}-\frac{1}{(1+w_{t})\rho_{t}}\frac{k^{2}}{a^{2}}\frac{Gf''}{\theta}\Big)\mathbf{G}\nonumber\\
&&-\frac{1}{(1+z)H}\Big[8H\left(f''b_{2}+f'''b_{1}^{2}+3Hf'-\frac{3}{2}f'''Hb_{1}\right)-\frac{2}{3}\theta-\big(\frac{1}{3}\theta^{2}+\frac{1}{2}(1+3w_{t})\rho_{t}\big)\frac{Gb_{1}}{\theta^{2}(1+w_{t})\rho_{t}}\nonumber\\&&+\frac{k^{2}}{a^{2}(1+w_{t})\rho_{t}}\frac{Gb_{1}}{\theta^{2}}\Big]Z
\label{eq30}\\
 \end{eqnarray}
\begin{eqnarray}
 && \mathcal{G}'= -\frac{1}{(1+z)H}\Big(1-\frac{b_{1} G f''}{(1+w_{t})\rho_{t} \theta}\Big)\mathsf{G} -\frac{1}{(1+z)H}\frac{b_{1}^{2} G}{(1+w_{t})\rho_{t} \theta^{2}} Z +\frac{1}{(1+z)H}\frac{b_{1} w_{m} \rho_{m} \Delta_{m}}{(1+w_{t})\rho_{t}} \nonumber\\&&-\frac{1}{(1+z)H}\frac{\alpha b_{1} \rho_{ch}w_{ch}}{(1+w_{t})\rho_{t}}\Delta_{ch}
 +\Big(\frac{1}{2(1+z)H}\frac{b_{1}}{(1+w_{t})\rho_{t}}\Big(f'+Gf''-1\Big)\nonumber\\&&+\frac{1}{2(1+z)H}\frac{b_{1}^{2}}{(1+w_{t})\rho_{t}}\Big(Gf''+\frac{f''}{\theta^{2}}\Big)\Big)\mathcal{G}
 \label{eq31}\\
 && \mathsf{G}'=-\frac{1}{(1+z)H}\Big(\frac{b_{3}}{b_{1}}-\frac{b_{1}{G}f''}{(1+w_{t})\rho_{t}\theta}\Big)\mathsf{G}+\Big[\frac{1}{(1+z)H}\frac{b_{2}b_{1}{G}}{(1+w_{t})\rho_{t}\theta^{2}}\Big]Z+\Big[\frac{1}{(1+z)H}\frac{b_{2}}{(1+w_{t})\rho_{t}}w_{m} \rho_{m}\Big] \Delta_{m}\nonumber\\&&-\Big[\frac{1}{(1+z)H}\frac{\alpha b_{2} \rho_{ch}w_{ch}}{(1+w_{t})\rho_{t}}\Big]\Delta_{ch}-\frac{1}{(1+z)H}\Big[\frac{b_{2}}{2(1+w_{t})\rho_{t}}(f'+Gf''-1)-\frac{b_{2}b_{1}}{(1+w_{t})\rho_{t}}(Gf''+\frac{f''}{\theta^{2}})\Big]\mathcal{G}
 \label{eq32}
 \end{eqnarray}
\begin{eqnarray}
&&\Delta'_{G}=-\frac{1}{(1+z)H}\Big[-\frac{4H\theta}{\rho_{G}}\Big(2H b_{1}f''' +(2c_{1}-2H^{2})f''\Big)\nonumber\\&&+\theta \Big(4H^{2}f''b_{2}+4H(2c_{1}-H^{2})f''b_{1}\Big)\Big(\frac{Gf''}{\theta(1+w_{t})\rho_{t}\rho_{G}}\Big)\Big]\mathsf{G}\nonumber\\&&
-\frac{1}{(1+z)H}\Big[4H\Big(H(b_{2}f'''+\frac{f''b_{2}}{b_{2}}+b_{2}^{2}f''')+(2c_{1}-H^{2})f''\Big)\nonumber\\&&+\theta\Big(4H^{2}(f''b_{2}+f'''b^{2}_{1}))+4H(2c_{1}-H^{2})f''b_{1}\Big(\frac{\frac{1}{2}+Gb_{1}f''+\frac{b_{1}f'''}{\theta^{2}}}{(1+w_{t})\rho_{t}\rho_{G}}\Big)\Big]\mathcal{G} +\nonumber\\
&&-\frac{1}{3((1+z)H)\rho_{G}}\Big[\Big(2H(f''b_{2}+f'''b_{1}^{2})-8Hf''b_{1}(\frac{G}{12H^{3}}+3H)\nonumber\\&&+(2c_{1}-H^{2})f''b_{1}-3\Big(4H^{2}(f''b_{2}+f'''b^{2}_{1})+4H(2c_{1}-H^{2})f''b_{1}\Big)\nonumber\\
&&+\theta(4H^{2}(f''b_{2}+f'''b^{2}_{1})-4H(2c_{1}-H^{2})f''b_{1})(\frac{Gb_{1}\theta^{2}}{(1+w_{t})\rho_{t}\rho_{G}})\Big)\nonumber\\ &&+3\theta(4H^{2}(f''b_{2}+f'''b^{2}_{1})+4H(2c_{1}-H^{2})f''b_{1})\Big(\frac{(1-f'-Gf'')Gb_{1}}{\theta^{2}(1+w_{t})\rho_{t}}\Big)\Big]Z
\nonumber\\
&&-\frac{3}{(1+z)\rho_{G}}\Big(4H^{2}(f''b_{2}+f'''b^{2}_{1})+4H(2c_{1}-H^{2})f''b_{1}\Big) \Delta_{G}\nonumber\\
&&-\frac{3}{(1+z)}\Big(-4H^{2}(f''b_{2}+f'''b^{2}_{1})+4H(2c_{1}-H^{2})f''b_{1}\Big)\frac{w_{m}\rho_{m}(1-f'-Gf'')}{(1+w_{t}\rho_{t}\rho_{G})}\Delta_{m}\nonumber\\
&&+\frac{3}{(1+z)} \Big(4H^{2}(f''b_{2}+f'''b^{2}_{1})+4H(2c_{1}-H^{2})f''b_{1}\Big)\Big(\frac{w_{ch}\rho_{ch}}{(1+w_{t})\rho_{t}\rho_{G}}\Big)w_{ch}(\alpha+1)\Delta_{chp}
\label{eq33}\\
 &&\Delta^{'}_{ch}= \frac{1}{(1+z)H}[1+\frac{Gb_{1}}{\theta(1+w_{t})\rho_{t}}](1+w_{ch})Z-\frac{1}{(1+z)H}\frac{\theta (1+w_{ch})}{(1+w_{t})\rho_{t}}w_{m}\rho_{m}\Delta_{m}\nonumber\\&&-\frac{3}{(1+z)}\Big[(1+\rho_{ch})\frac{ (1+w_{ch})}{(1+w_{t})\rho_{t}}w_{ch}\rho_{ch}-(1+\alpha)w_{ch}\Big]\Delta_{ch}\nonumber\\&&-\frac{\theta (1+w_{ch})}{(1+z)(1+w_{t})H\rho_{t}}[\frac{1}{2}(1-f'-Gf'')+Gb_{1}f''+\frac{b_{1}f'''}{\theta^{2}}]\mathcal{G}\nonumber\\
 &&-\frac{\theta (1+w_{ch})Gf''}{(1+z)(1+w_{t})H\rho_{t}}\mathsf{G}\;,
 \label{eq34}
\end{eqnarray}
 with $k=\frac{2\pi a}{\lambda}$, $k$ being the wave number and $\lambda$, the wavelength of perturbations. The parameters $b_{1}$, $b_{2}$, $b_{3}$ and  $c_{1}$ are presented in Appendix. According to eq. \ref{eq7}, the vector term $a \tilde{\bigtriangledown}_{a}w_{ch}$ presents its scalar part as $a \tilde{\bigtriangledown}_{a}w_{ch}=-w_{ch}(1+\alpha)\Delta_{ch}$. In GR limits, with normal form of matter, one can obtain a closed system of first-order perturbation equations which is easier to find the analytical solutions. However, the linear perturbation equations are coupled system of first -order  equations for the density fluctuations of matter, chaplygin and Gauss-Bonnet fluids which are more complicated to find the analytical solutions. We therefore considered short wavelength($\frac{k^{2}}{a^{2}H^{2}}\gg1$) and long wavelength($\frac{k^{2}}{a^{2}H^{2}}\ll1$) limits of the perturbation to analyse the large scale structure implications of the numerical results in the redshift space.
\section{Asymptotic Analysis}\label{section4}
In this section , we analyse the evolution of the perturbation equations in both long-wavelength and short-wavelength regimes together with the consideration that the universe is  dominated  by dust.
\subsection{Matter density fluctuations in GR limits}
\begin{figure}
  \includegraphics[width=120mm,height=85mm]{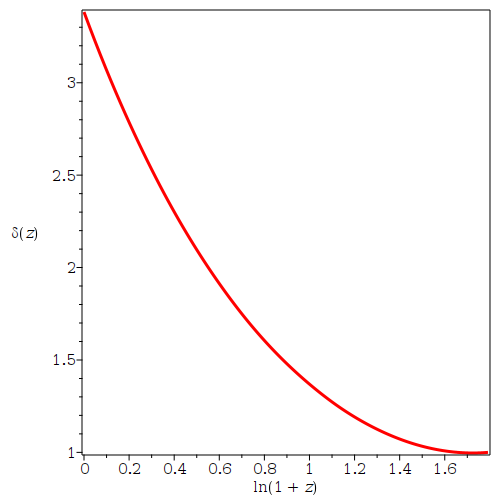}
  \caption{Plot of matter energy overdensity perturbations versus redshift of Eq. \ref{eq35} using eq. \ref{eq36}-\ref{eq37} for GR limit ($f(G)=G$) for a dust($w=0$) dominated Universe for $\Delta_{in}=10^{-5}$, $Z_{in}=10^{-5}$.}
  \label{Fig1}
 \end{figure}

In this part, we analyse the behavior of energy overdensity fluctuations for dust fluid in GR limits for both long and short wavelength modes for the case $f(G)=G$ and no contribution from the chaplygin gas fluid. We also define the normalised energy density contrast for energy density fluid as
\begin{equation*}
 \delta(z)=\frac{\Delta^{k}_{m}(z)}{\Delta(z_{0})},
\end{equation*}
where $\Delta(z_{0})$ is the matter energy density at the initial redshift, hereafter $z_{0}=1100$.
 If we assume that the Universe is dominated mainly by dust fluid, the equation of state parameter becomes $w=0$. Consequently eq. \ref{eq29} through to eq. \ref{eq34} become
\begin{eqnarray}
 &&\Delta'_{m}=\frac{1}{(1+z)H}Z,
 \label{eq35}\\
 && Z'=\frac{\rho_{d}}{2(1+z)H},
 \label{eq36}\\
&& \mathcal{G}'=0,
 \mathsf{G}'=0,
  \Delta'_{G}=0,
  \Delta'_{ch}=0,
  \label{eq37}
  \end{eqnarray}
   which admit the numerical solutions presented in fig. \ref{Fig1} for  dust dominated universe. The matter energy overdensity fluctuations $\delta(z)$ decay with increase in redshift. Throughout all the plots, we rescaled the $\delta(z)$ to make it readable.
\subsection{ Energy overdensity fluctuations in the context of $f(G)$ gravity approach}
For pedagogical purpose, we consider a polynomial $f(G)$ model  presents in \cite{venikoudis2022late} given by
\begin{equation}
 f(G)=\frac{\alpha_{1}}{G^{2}}+\alpha_{2}G^{\frac{1}{2}},
 \label{eq64}
\end{equation}
   We choose this $f(G)$ model for a quantitative analysis of the derived perturbation equations eq. \ref{eq29} through to eq. \ref{eq34}, it is a  viable model which is  compatible with cosmological observations and  it is a viable model accounts for the late-time acceleration of the universe without the need for dark energy.  The Hubble parameter in redshift space  for a dust dominated universe ($w=0$) is presented as $H(z)=\frac{2m}{3}(1+z)^{\frac{3}{2m}}$, for GR limits, we set $m=1$. The Gauss-Bonnet term is given by $ G=24H^{2}\left(H^{2}+\dot{H}\right)$ which in redshift space is presented as $G=\frac{64}{9}m^{3}\Big(\frac{2m}{3}-1\Big)(1+z)^{\frac{6}{m}}$.  We assume that the dynamics of the universe is driven by  the power-law scale factor of the form $a(t)=t^{\frac{2m}{3(1+w)}}$ and we consider $H=\frac{\dot{a}}{a}$. We analyse the perturbation equations in both short- and long- wavelength modes.
\subsubsection{Short-Wavelength mode}
\begin{figure}
  \includegraphics[width=130mm,height=80mm]{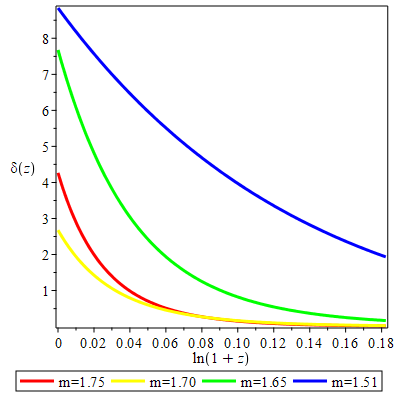}
  \caption{Plot of energy density perturbations versus redshift of Eq. \ref{eq39}-\ref{eq44} for different values of $m$ in the dust dominated Universe for a short wavelength mode $k=100$ and $\alpha=0.2$.}
  \label{Fig2}
 \end{figure}

\begin{figure}
  \includegraphics[width=120mm,height=80mm]{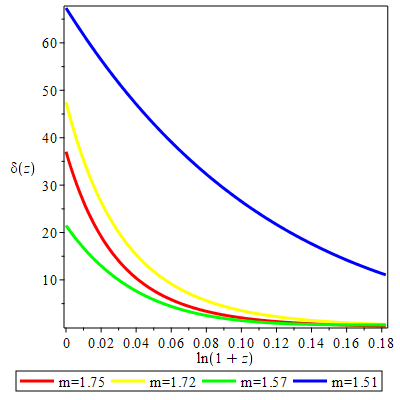}
  \caption{Plot of energy density perturbations versus redshift of Eq. \ref{eq45}-\ref{eq50} for different values of $m$ in the dust dominated Universe for short wavelength mode $k=1000$ in a generalized Chaplygin model $\alpha=1$.}
  \label{Fig3}
 \end{figure}

 In this subsection, we first consider the short-wavelength regime
  where  $\frac{k^{2}}{a^{2}}>>1$, with $k=\frac{2\pi a}{\lambda}$ and $\lambda$ is the wavelength of the mode. Applying this approximation and considering the dust dominated universe, it means $w_{m}=0$, eq. \ref{eq29} through to eq. \ref{eq34} are represented as
\begin{eqnarray}
 &&\Delta'_{m}=\frac{1}{(1+z)H}\Big(\Big(1-\frac{1}{(1+w_{t})\rho_{t}}\Big)\frac{(1+z)HGG'}{\theta^{2}}\Big)Z\nonumber\\&&+\frac{1}{(1+z)H}\alpha \frac{1}{(1+w_{t})\rho_{t}}w_{chp}\rho_{chp}\Delta_{chp}\nonumber\\&&+\frac{1}{(1+z)H}\frac{1}{(1+w_{t})\rho_{t}}\frac{Gf''}{\theta}\mathsf{G}
-\frac{1}{(1+z)H}\frac{1}{(1+w_{t})\rho_{t}}\Big(\frac{1}{2}(1-f'-Gf'')G(1+z)HG'f'''\nonumber\\&&-\frac{(1+z)HG'f''}{G^{2}}\Big)\mathcal{G}
 \label{eq39}
\end{eqnarray}
\begin{eqnarray}
&& Z'=\frac{1}{(1+z)H}\Big(\frac{1}{2}\rho_{d}+\frac{1}{3}\theta^{2}\Big)\Delta_{m}\nonumber\\
&&-\frac{1}{(1+z)H}\Big[-\frac{1}{2}(1+3w_{chp})\rho_{chp}+\big(\frac{1}{3}\theta^{2}+\frac{1}{2}(1+3w_{t})\rho_{t}\big)\frac{w_{chp}\rho_{chp}}{(1+w_{t})\rho_{t}}\nonumber\\
&&\Big[\frac{3}{2}\rho_{chp}-\big(\frac{1}{3}\theta^{2}+\frac{1}{2}(1+3w_{t})\rho_{t}\big)\frac{\rho_{chp}}{(1+w_{t})\rho_{t}}\Big](\alpha+1)w_{chp}-\frac{\big(w_{chp}\rho_{chp}-(\alpha+1)w_{chp}\big)}{(1+w_{t})\rho_{t}}\frac{k^{2}}{a^{2}}\Big]\Delta_{chp}\nonumber\\&&-\frac{1}{(1+z)H}\Big[1-3f''-12H^{3}b_{1}f^{iv}+12H\Big(H(b_{2}f'''+\frac{b_{3}}{b_{1}}f''+b_{1}^{2}f^{iv})+2H^{3}f''\Big)\nonumber\\
&&+\big(\frac{1}{3}\theta^{2}+\frac{1}{2}(1+3w_{t})\rho_{t}\big)\frac{1}{(1+w_{t})\rho_{t}}\Big(\frac{1}{2}(1-f'-Gf'-Gf'')+Gb_{1}f'''+\frac{b_{1}f''}{G^{2}}\Big)\nonumber\\&& -\frac{1}{(1+w_{t})\rho_{t}}\frac{k^{2}}{a^{2}}\Big(\frac{1}{2}(1-f'-Gf'')+Gb_{1}f'''+\frac{b_{1}f''}{G^{2}}\Big)\Big]\mathcal{G}\nonumber\\
&&-\frac{1}{(1+z)H}\Big(24b_{1}f'''-12H^{3}f'''+\big(\frac{1}{3}\theta^{2}+\frac{1}{2}(1+3w_{t})\rho_{t}\big)\frac{Gf''}{\theta(1+w_{t})\rho_{t}}-\frac{1}{(1+w_{t})\rho_{t}}\frac{k^{2}}{a^{2}}\frac{Gf''}{\theta}\Big)\mathbf{G}\nonumber\\
&&-\frac{1}{(1+z)H}\Big[8H\left(f''b_{2}+f'''b_{1}^{2}+3Hf'-\frac{3}{2}f'''Hb_{1}\right)-\frac{2}{3}\theta-\big(\frac{1}{3}\theta^{2}+\frac{1}{2}(1+3w_{t})\rho_{t}\big)\frac{Gb_{1}}{\theta^{2}(1+w_{t})\rho_{t}}\nonumber\\&&+\frac{k^{2}}{a^{2}(1+w_{t})\rho_{t}}\frac{Gb_{1}}{\theta^{2}}\Big]Z
\label{eq40}\\
\end{eqnarray}
\begin{eqnarray}
 && \mathcal{G}'= -\frac{1}{(1+z)H}\Big(1-\frac{b_{1} G f''}{(1+w_{t})\rho_{t} \theta}\Big)\mathsf{G} -\frac{1}{(1+z)H}\frac{b_{1}^{2} G}{(1+w_{t})\rho_{t} \theta^{2}} Z  \nonumber\\&&-\frac{1}{(1+z)H}\frac{\alpha b_{1} \rho_{ch}w_{ch}}{(1+w_{t})\rho_{t}}\Delta_{ch}
 +\Big(\frac{1}{2(1+z)H}\frac{b_{1}}{(1+w_{t})\rho_{t}}\Big(f'+Gf''-1\Big)\nonumber\\&&+\frac{1}{2(1+z)H}\frac{b_{1}^{2}}{(1+w_{t})\rho_{t}}\Big(Gf''+\frac{f''}{\theta^{2}}\Big)\Big)\mathcal{G}
 \label{eq41}\\
 && \mathsf{G}'=-\frac{1}{(1+z)H}\Big(\frac{b_{3}}{b_{1}}-\frac{b_{1}{G}f''}{(1+w_{t})\rho_{t}\theta}\Big)\mathsf{G}+\Big[\frac{1}{(1+z)H}\frac{b_{2}b_{1}{G}}{(1+w_{t})\rho_{t}\theta^{2}}\Big]Z-\Big[\frac{1}{(1+z)H}\frac{\alpha b_{2} \rho_{ch}w_{ch}}{(1+w_{t})\rho_{t}}\Big]\Delta_{ch}\nonumber\\&&-\frac{1}{(1+z)H}\Big[\frac{b_{2}}{2(1+w_{t})\rho_{t}}(f'+Gf''-1)-\frac{b_{2}b_{1}}{(1+w_{t})\rho_{t}}(Gf''+\frac{f''}{\theta^{2}})\Big]\mathcal{G}
 \label{eq42}
 \end{eqnarray}
\begin{eqnarray}
&&\Delta'_{G}=-\frac{1}{(1+z)H}\Big[-\frac{4H\theta}{\rho_{G}}\Big[2H b_{1}f''' +(2c_{1}-2H^{2})f''\Big]\nonumber\\&&+\theta \Big(4H^{2}f''b_{2}+4H(2c_{1}-H^{2})f''b_{1}\Big)\Big(\frac{Gf''}{\theta(1+w_{t})\rho_{t}\rho_{G}}\Big)\Big]\mathsf{G}\nonumber\\&&
-\frac{1}{(1+z)H}\Big[4H\Big[H(b_{2}f'''+\frac{f''b_{2}}{b_{2}}+b_{2}^{2}f''')+(2c_{1}-H^{2})f''\Big]\nonumber\\&&+\theta\Big(4H^{2}(f''b_{2}+f'''b^{2}_{1}))+4H(2c_{1}-H^{2})f''b_{1}\Big(\frac{\frac{1}{2}+Gb_{1}f''+\frac{b_{1}f'''}{\theta^{2}}}{(1+w_{t})\rho_{t}\rho_{G}}\Big)\Big]\mathcal{G} +\nonumber\\
&&-\frac{1}{3((1+z)H)\rho_{G}}\Big[\Big(2H(f''b_{2}+f'''b_{1}^{2})-8Hf''b_{1}(\frac{G}{12H^{3}}+3H)\nonumber\\&&+(2c_{1}-H^{2})f''b_{1}-3\Big(4H^{2}(f''b_{2}+f'''b^{2}_{1})+4H(2c_{1}-H^{2})f''b_{1}\Big)\nonumber\\
&&+\theta(4H^{2}(f''b_{2}+f'''b^{2}_{1})-4H(2c_{1}-H^{2})f''b_{1})(\frac{Gb_{1}\theta^{2}}{(1+w_{t})\rho_{t}\rho_{G}})\Big)\nonumber\\ &&+3\theta(4H^{2}(f''b_{2}+f'''b^{2}_{1})+4H(2c_{1}-H^{2})f''b_{1})\Big(\frac{(1-f'-Gf'')Gb_{1}}{\theta^{2}(1+w_{t})\rho_{t}}\Big)\Big]Z
\nonumber\\
&&-\frac{3}{(1+z)\rho_{G}}\Big(4H^{2}(f''b_{2}+f'''b^{2}_{1})+4H(2c_{1}-H^{2})f''b_{1}\Big) \Delta_{G}\nonumber\\&&+\frac{3}{(1+z)} \Big(4H^{2}(f''b_{2}+f'''b^{2}_{1})+4H(2c_{1}-H^{2})f''b_{1}\Big)\Big(\frac{w_{ch}\rho_{ch}}{(1+w_{t})\rho_{t}\rho_{G}}\Big)w_{ch}(\alpha+1)\Delta_{chp}
\label{eq43}\\
 &&\Delta^{'}_{ch}= \frac{1}{(1+z)H}[1+\frac{Gb_{1}}{\theta(1+w_{t})\rho_{t}}](1+w_{ch})Z\nonumber\\&&-\frac{3}{(1+z)}\Big[(1+\rho_{ch})\frac{ (1+w_{ch})}{(1+w_{t})\rho_{t}}w_{ch}\rho_{ch}-(1+\alpha)w_{ch}\Big]\Delta_{ch}-\nonumber\\&&\frac{\theta (1+w_{ch})}{(1+z)(1+w_{t})H\rho_{t}}[\frac{1}{2}(1-f'-Gf'')+Gb_{1}f''+\frac{b_{1}f'''}{\theta^{2}}]\mathcal{G}-\frac{\theta (1+w_{ch})Gf''}{(1+z)(1+w_{t})H\rho_{t}}\mathsf{G}\;
 \label{eq44}
\end{eqnarray}
The energy density for the chaplygin gas fluid $\Delta_{ch}$ and the energy density for Gauss-Bonnet fluid $\Delta_{G}$ do not couple with the matter energy density $\Delta_{m}$.
Numerical solutions of eq. \ref{eq39} through to eq. \ref{eq44} are  presented in fig. \ref{Fig2} for different values of parameter $m$.
 For simplicity we set $f(G)$ to $f$  and considered the initial conditions $\mathcal{G}_{in}=10^{-5}$, $\mathbf{G}_{in}=10^{-5}$ and $\Delta_{in}=10^{-5}$   to find numerical solutions.
  The equations above are presented for the generalized Chaplygin gas model. During numerical results computation, where there is $w_{t}$ and $\rho_{t}$ terms, we replaced these terms by relations of eq. \ref{eq3}-\ref{eq7} and eq. \ref{eq12}-\ref{eq13} throughout all equations. The original Chaplygin gas  model can be obtained by setting $\alpha=1$  in eq. \ref{eq39} through to eq. \ref{eq44} and  its evolution equations are presented as \begin{eqnarray}
 &&\Delta'_{m}=\frac{1}{(1+z)H}\Big(\Big(1-\frac{1}{(1+w_{t})\rho_{t}}\Big)\frac{(1+z)HGG'}{\theta^{2}}\Big)Z+\frac{1}{(1+z)H} \frac{1}{(1+w_{t})\rho_{t}}w_{chp}\rho_{chp}\Delta_{chp}\nonumber\\&&+\frac{1}{(1+z)H}\frac{1}{(1+w_{t})\rho_{t}}\frac{Gf''}{\theta}\mathsf{G}
-\frac{1}{(1+z)H}\frac{1}{(1+w_{t})\rho_{t}}\Big(\frac{1}{2}(1-f'-Gf'')G(1+z)HG'f'''\nonumber\\&&-\frac{(1+z)HG'f''}{G^{2}}\Big)\mathcal{G}
 \label{eq45}
 \end{eqnarray}
 \begin{eqnarray}
&& Z'=\frac{1}{(1+z)H}\Big(\frac{1}{2}\rho_{d}+\frac{1}{3}\theta^{2}\Big)\Delta_{m}\nonumber\\
&&-\frac{1}{(1+z)H}\Big[-\frac{1}{2}(1+3w_{chp})\rho_{chp}+\big(\frac{1}{3}\theta^{2}+\frac{1}{2}(1+3w_{t})\rho_{t}\big)\frac{w_{chp}\rho_{chp}}{(1+w_{t})\rho_{t}}\nonumber\\
&&2\Big[\frac{3}{2}\rho_{chp}-\big(\frac{1}{3}\theta^{2}+\frac{1}{2}(1+3w_{t})\rho_{t}\big)\frac{\rho_{chp}}{(1+w_{t})\rho_{t}}\Big]w_{chp}-\frac{\big(w_{chp}\rho_{chp}-2w_{chp}\big)}{(1+w_{t})\rho_{t}}\frac{k^{2}}{a^{2}}\Big]\Delta_{ch}\nonumber\\&&-\frac{1}{(1+z)H}\Big[1-3f''-12H^{3}b_{1}f^{iv}+12H\Big(H(b_{2}f'''+\frac{b_{3}}{b_{1}}f''+b_{1}^{2}f^{iv})+2H^{3}f''\Big)\nonumber\\
&&+\big(\frac{1}{3}\theta^{2}+\frac{1}{2}(1+3w_{t})\rho_{t}\big)\frac{1}{(1+w_{t})\rho_{t}}\Big(\frac{1}{2}(1-f'-Gf'-Gf'')+Gb_{1}f'''+\frac{b_{1}f''}{G^{2}}\Big)\nonumber\\&& -\frac{1}{(1+w_{t})\rho_{t}}\frac{k^{2}}{a^{2}}\Big(\frac{1}{2}(1-f'-Gf'')+Gb_{1}f'''+\frac{b_{1}f''}{G^{2}}\Big)\Big]\mathcal{G}\nonumber\\
&&-\frac{1}{(1+z)H}\Big(24b_{1}f'''-12H^{3}f'''+\big(\frac{1}{3}\theta^{2}+\frac{1}{2}(1+3w_{t})\rho_{t}\big)\frac{Gf''}{\theta(1+w_{t})\rho_{t}}-\frac{1}{(1+w_{t})\rho_{t}}\frac{k^{2}}{a^{2}}\frac{Gf''}{\theta}\Big)\mathbf{G}\nonumber\\
&&-\frac{1}{(1+z)H}\Big[8H\left(f''b_{2}+f'''b_{1}^{2}+3Hf'-\frac{3}{2}f'''Hb_{1}\right)-\frac{2}{3}\theta-\big(\frac{1}{3}\theta^{2}+\frac{1}{2}(1+3w_{t})\rho_{t}\big)\frac{Gb_{1}}{\theta^{2}(1+w_{t})\rho_{t}}\nonumber\\&&+\frac{k^{2}}{a^{2}(1+w_{t})\rho_{t}}\frac{Gb_{1}}{\theta^{2}}\Big]Z
\label{eq46}\\
 && \mathcal{G}'= -\frac{1}{(1+z)H}\Big(1-\frac{b_{1} G f''}{(1+w_{t})\rho_{t} \theta}\Big)\mathsf{G} -\frac{1}{(1+z)H}\frac{b_{1}^{2} G}{(1+w_{t})\rho_{t} \theta^{2}} Z  \nonumber\\&&-\frac{1}{(1+z)H}\frac{ b_{1} \rho_{ch}w_{ch}}{(1+w_{t})\rho_{t}}\Delta_{ch}
 +\Big(\frac{1}{2(1+z)H}\frac{b_{1}}{(1+w_{t})\rho_{t}}\Big(f'+Gf''-1\Big)\nonumber\\&&+\frac{1}{2(1+z)H}\frac{b_{1}^{2}}{(1+w_{t})\rho_{t}}\Big(Gf''+\frac{f''}{\theta^{2}}\Big)\Big)\mathcal{G}
 \label{eq47}\\
 && \mathsf{G}'=-\frac{1}{(1+z)H}\Big(\frac{b_{3}}{b_{1}}-\frac{b_{1}{G}f''}{(1+w_{t})\rho_{t}\theta}\Big)\mathsf{G}+\Big[\frac{1}{(1+z)H}\frac{b_{2}b_{1}{G}}{(1+w_{t})\rho_{t}\theta^{2}}\Big]Z-\Big[\frac{1}{(1+z)H}\frac{ b_{2} \rho_{ch}w_{ch}}{(1+w_{t})\rho_{t}}\Big]\Delta_{ch}\nonumber\\&&-\frac{1}{(1+z)H}\Big[\frac{b_{2}}{2(1+w_{t})\rho_{t}}(f'+Gf''-1)-\frac{b_{2}b_{1}}{(1+w_{t})\rho_{t}}(Gf''+\frac{f''}{\theta^{2}})\Big]\mathcal{G}
 \label{eq48}
 \end{eqnarray}
\begin{eqnarray}
&&\Delta'_{G}=-\frac{1}{(1+z)H}\Big[-\frac{4H\theta}{\rho_{G}}\Big[2H b_{1}f''' +(2c_{1}-2H^{2})f''\Big]\nonumber\\&&+\theta \Big(4H^{2}f''b_{2}+4H(2c_{1}-H^{2})f''b_{1}\Big)\Big(\frac{Gf''}{\theta(1+w_{t})\rho_{t}\rho_{G}}\Big)\Big]\mathsf{G}\nonumber\\&&
-\frac{1}{(1+z)H}\Big[4H\Big[H(b_{2}f'''+\frac{f''b_{2}}{b_{2}}+b_{2}^{2}f''')+(2c_{1}-H^{2})f''\Big]\nonumber\\&&+\theta\Big(4H^{2}(f''b_{2}+f'''b^{2}_{1}))+4H(2c_{1}-H^{2})f''b_{1}\Big(\frac{\frac{1}{2}+Gb_{1}f''+\frac{b_{1}f'''}{\theta^{2}}}{(1+w_{t})\rho_{t}\rho_{G}}\Big)\Big]\mathcal{G} +\nonumber\\
&&-\frac{1}{3((1+z)H)\rho_{G}}\Big[\Big(2H(f''b_{2}+f'''b_{1}^{2})-8Hf''b_{1}(\frac{G}{12H^{3}}+3H)\nonumber\\&&+(2c_{1}-H^{2})f''b_{1}-3\Big(4H^{2}(f''b_{2}+f'''b^{2}_{1})+4H(2c_{1}-H^{2})f''b_{1}\Big)\nonumber\\
&&+\theta(4H^{2}(f''b_{2}+f'''b^{2}_{1})-4H(2c_{1}-H^{2})f''b_{1})(\frac{Gb_{1}\theta^{2}}{(1+w_{t})\rho_{t}\rho_{G}})\Big)\nonumber\\ &&+3\theta(4H^{2}(f''b_{2}+f'''b^{2}_{1})+4H(2c_{1}-H^{2})f''b_{1})\Big(\frac{(1-f'-Gf'')Gb_{1}}{\theta^{2}(1+w_{t})\rho_{t}}\Big)\Big]Z
\nonumber\\
&&-\frac{3}{(1+z)\rho_{G}}\Big(4H^{2}(f''b_{2}+f'''b^{2}_{1})+4H(2c_{1}-H^{2})f''b_{1}\Big) \Delta_{G}\nonumber\\&&+\frac{6}{(1+z)} \Big(4H^{2}(f''b_{2}+f'''b^{2}_{1})+4H(2c_{1}-H^{2})f''b_{1}\Big)\Big(\frac{w_{ch}\rho_{ch}}{(1+w_{t})\rho_{t}\rho_{G}}\Big)w_{ch}\Delta_{chp}
\label{eq49}\\
 &&\Delta^{'}_{ch}= \frac{1}{(1+z)H}[1+\frac{Gb_{1}}{\theta(1+w_{t})\rho_{t}}](1+w_{ch})Z\nonumber\\&&-\frac{3}{(1+z)}\Big[(1+\rho_{ch})\frac{ (1+w_{ch})}{(1+w_{t})\rho_{t}}w_{ch}\rho_{ch}-(1+\alpha)w_{ch}\Big]\Delta_{ch}\nonumber\\&&-\frac{\theta (1+w_{ch})}{(1+z)(1+w_{t})H\rho_{t}}[\frac{1}{2}(1-f'-Gf'')+Gb_{1}f''+\frac{b_{1}f'''}{\theta^{2}}]\mathcal{G}\nonumber\\&&-\frac{\theta (1+w_{ch})Gf''}{(1+z)(1+w_{t})H\rho_{t}}\mathsf{G}
 \label{eq50}
\end{eqnarray}
which admit numerical solutions presented in fig. \ref{Fig3}.
We considered the initial conditions $\mathcal{G}_{in}=10^{-5}$, $\mathbf{G}_{in}=10^{-5}$ and $\Delta_{in}=10^{-5}$   to find numerical solutions of Eq. \ref{eq45} through to Eq. \ref{eq50}. From the plot, the energy overdensity decay with increase in redshift.
 \subsubsection{Long-Wavelength mode}
  In this subsection, we consider the long-wavelength regime
  where  $\frac{k^{2}}{a^{2}}<< 1$, this means that  all $k$'s terms  become negligible. Applying this approximation, and consider  the dust dominated Universe,  eq.\ref{eq29} through to eq. \ref{eq34} are represented as
\begin{eqnarray}
 &&\Delta'_{m}=\frac{1}{(1+z)H}\Big(\Big(1-\frac{1}{(1+w_{t})\rho_{t}}\Big)\frac{(1+z)HGG'}{\theta^{2}}\Big)Z\nonumber\\&&+\frac{1}{(1+z)H}\alpha \frac{1}{(1+w_{t})\rho_{t}}w_{chp}\rho_{chp}\Delta_{chp}\nonumber\\&&+\frac{1}{(1+z)H}\frac{1}{(1+w_{t})\rho_{t}}\frac{Gf''}{\theta}\mathsf{G}
-\frac{1}{(1+z)H}\frac{1}{(1+w_{t})\rho_{t}}\Big(\frac{1}{2}(1-f'-Gf'')G(1+z)HG'f'''\nonumber\\&&-\frac{(1+z)HG'f''}{G^{2}}\Big)\mathcal{G}
 \label{eq51}
\end{eqnarray}
\begin{eqnarray}
&& Z'=\frac{1}{(1+z)H}\Big(\frac{1}{2}\rho_{d}+\frac{1}{3}\theta^{2}\Big)\Delta_{m}\nonumber\\
&&-\frac{1}{(1+z)H}\Big[-\frac{1}{2}(1+3w_{chp})\rho_{chp}+\big(\frac{1}{3}\theta^{2}+\frac{1}{2}(1+3w_{t})\rho_{t}\big)\frac{w_{chp}\rho_{chp}}{(1+w_{t})\rho_{t}}\nonumber\\
&&\Big[\frac{3}{2}\rho_{chp}-\big(\frac{1}{3}\theta^{2}+\frac{1}{2}(1+3w_{t})\rho_{t}\big)\frac{\rho_{chp}}{(1+w_{t})\rho_{t}}\Big](\alpha+1)w_{chp}\Big]\Delta_{chp}\nonumber\\&&-\frac{1}{(1+z)H}\Big[1-3f''-12H^{3}b_{1}f^{iv}+12H\Big(H(b_{2}f'''+\frac{b_{3}}{b_{1}}f''+b_{1}^{2}f^{iv})+2H^{3}f''\Big)\nonumber\\
&&+\big(\frac{1}{3}\theta^{2}+\frac{1}{2}(1+3w_{t})\rho_{t}\big)\frac{1}{(1+w_{t})\rho_{t}}\Big(\frac{1}{2}(1-f'-Gf'-Gf'')+Gb_{1}f'''+\frac{b_{1}f''}{G^{2}}\Big)\Big]\mathcal{G}\nonumber\\
&&-\frac{1}{(1+z)H}\Big(24b_{1}f'''-12H^{3}f'''+\big(\frac{1}{3}\theta^{2}+\frac{1}{2}(1+3w_{t})\rho_{t}\big)\frac{Gf''}{\theta(1+w_{t})\rho_{t}}\Big)\mathbf{G}\nonumber\\
&&-\frac{1}{(1+z)H}\Big[8H\left(f''b_{2}+f'''b_{1}^{2}+3Hf'-\frac{3}{2}f'''Hb_{1}\right)-\frac{2}{3}\theta\nonumber\\&&-\big(\frac{1}{3}\theta^{2}+\frac{1}{2}(1+3w_{t})\rho_{t}\big)\frac{Gb_{1}}{\theta^{2}(1+w_{t})\rho_{t}}\Big]Z
\label{eq52}\\
 && \mathcal{G}'= -\frac{1}{(1+z)H}\Big(1-\frac{b_{1} G f''}{(1+w_{t})\rho_{t} \theta}\Big)\mathsf{G} -\frac{1}{(1+z)H}\frac{b_{1}^{2} G}{(1+w_{t})\rho_{t} \theta^{2}} Z  \nonumber\\&&-\frac{1}{(1+z)H}\frac{\alpha b_{1} \rho_{ch}w_{ch}}{(1+w_{t})\rho_{t}}\Delta_{ch}
 +\Big(\frac{1}{2(1+z)H}\frac{b_{1}}{(1+w_{t})\rho_{t}}\Big(f'+Gf''-1\Big)\nonumber\\&&+\frac{1}{2(1+z)H}\frac{b_{1}^{2}}{(1+w_{t})\rho_{t}}\Big(Gf''+\frac{f''}{\theta^{2}}\Big)\Big)\mathcal{G}
 \label{eq53}\\
 && \mathsf{G}'=-\frac{1}{(1+z)H}\Big(\frac{b_{3}}{b_{1}}-\frac{b_{1}{G}f''}{(1+w_{t})\rho_{t}\theta}\Big)\mathsf{G}+\Big[\frac{1}{(1+z)H}\frac{b_{2}b_{1}{G}}{(1+w_{t})\rho_{t}\theta^{2}}\Big]Z-\Big[\frac{1}{(1+z)H}\frac{\alpha b_{2} \rho_{ch}w_{ch}}{(1+w_{t})\rho_{t}}\Big]\Delta_{ch}\nonumber\\&&-\frac{1}{(1+z)H}\Big[\frac{b_{2}}{2(1+w_{t})\rho_{t}}(f'+Gf''-1)-\frac{b_{2}b_{1}}{(1+w_{t})\rho_{t}}(Gf''+\frac{f''}{\theta^{2}})\Big]\mathcal{G}
 \label{eq54}
 \end{eqnarray}
\begin{eqnarray}
&&\Delta'_{G}=-\frac{1}{(1+z)H}\Big[-\frac{4H\theta}{\rho_{G}}\Big[2H b_{1}f''' +(2c_{1}-2H^{2})f''\Big]\nonumber\\&&+\theta \Big(4H^{2}f''b_{2}+4H(2c_{1}-H^{2})f''b_{1}\Big)\Big(\frac{Gf''}{\theta(1+w_{t})\rho_{t}\rho_{G}}\Big)\Big]\mathsf{G}\nonumber\\&&
-\frac{1}{(1+z)H}\Big[4H\Big[H(b_{2}f'''+\frac{f''b_{2}}{b_{2}}+b_{2}^{2}f''')+(2c_{1}-H^{2})f''\Big]\nonumber\\&&+\theta\Big(4H^{2}(f''b_{2}+f'''b^{2}_{1}))+4H(2c_{1}-H^{2})f''b_{1}\Big(\frac{\frac{1}{2}+Gb_{1}f''+\frac{b_{1}f'''}{\theta^{2}}}{(1+w_{t})\rho_{t}\rho_{G}}\Big)\Big]\mathcal{G} +\nonumber\\
&&-\frac{1}{3((1+z)H)\rho_{G}}\Big[\Big(2H(f''b_{2}+f'''b_{1}^{2})-8Hf''b_{1}(\frac{G}{12H^{3}}+3H)\nonumber\\&&+(2c_{1}-H^{2})f''b_{1}-3\Big(4H^{2}(f''b_{2}+f'''b^{2}_{1})+4H(2c_{1}-H^{2})f''b_{1}\Big)\nonumber\\
&&+\theta(4H^{2}(f''b_{2}+f'''b^{2}_{1})-4H(2c_{1}-H^{2})f''b_{1})(\frac{Gb_{1}\theta^{2}}{(1+w_{t})\rho_{t}\rho_{G}})\Big)\nonumber\\ &&+3\theta(4H^{2}(f''b_{2}+f'''b^{2}_{1})+4H(2c_{1}-H^{2})f''b_{1})\Big(\frac{(1-f'-Gf'')Gb_{1}}{\theta^{2}(1+w_{t})\rho_{t}}\Big)\Big]Z
\nonumber\\
&&-\frac{3}{(1+z)\rho_{G}}\Big(4H^{2}(f''b_{2}+f'''b^{2}_{1})+4H(2c_{1}-H^{2})f''b_{1}\Big) \Delta_{G}\nonumber\\&&+\frac{3}{(1+z)} \Big(4H^{2}(f''b_{2}+f'''b^{2}_{1})+4H(2c_{1}-H^{2})f''b_{1}\Big)\Big(\frac{w_{ch}\rho_{ch}}{(1+w_{t})\rho_{t}\rho_{G}}\Big)w_{ch}(\alpha+1)\Delta_{chp}
\label{eq55}\\
 &&\Delta^{'}_{ch}= \frac{1}{(1+z)H}[1+\frac{Gb_{1}}{\theta(1+w_{t})\rho_{t}}](1+w_{ch})Z\nonumber\\&&-\frac{3}{(1+z)}\Big[(1+\rho_{ch})\frac{ (1+w_{ch})}{(1+w_{t})\rho_{t}}w_{ch}\rho_{ch}-(1+\alpha)w_{ch}\Big]\Delta_{ch}-\nonumber\\&&\frac{\theta (1+w_{ch})}{(1+z)(1+w_{t})H\rho_{t}}[\frac{1}{2}(1-f'-Gf'')+Gb_{1}f''+\frac{b_{1}f'''}{\theta^{2}}]\mathcal{G}-\frac{\theta (1+w_{ch})Gf''}{(1+z)(1+w_{t})H\rho_{t}}\mathsf{G}
 \label{eq56}
\end{eqnarray}
\begin{figure}[H]
  \includegraphics[width=130mm,height=90mm]{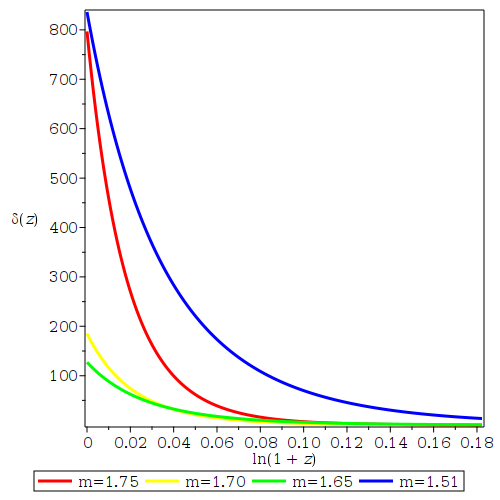}
  \caption{Plot of energy density perturbations versus redshift of Eq. \ref{eq51}-\ref{eq56} for different values of $m$ in the dust dominated Universe for long wavelength mode $k=0$ and $\alpha=0.2$.}
  \label{Fig4}
\end{figure}
 \begin{figure}[H]
  \includegraphics[width=130mm,height=90mm]{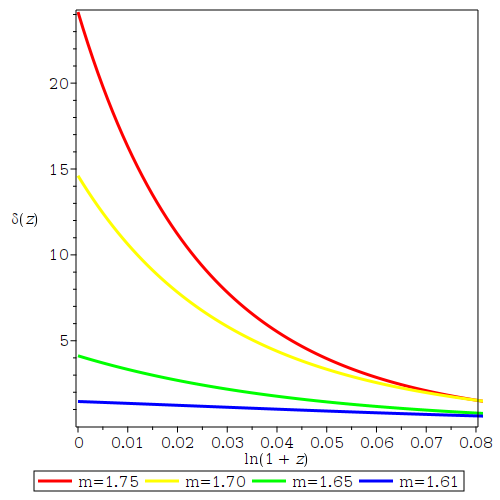}
  \caption{Plot of energy density perturbations versus redshift of Eq. \ref{eq51}-\ref{eq56} for different values of $m$ in the dust dominated Universe for long wavelength mode $k=0.001$ and $\alpha=0.2$.}
  \label{Fig5}
 \end{figure}
  For simplicity we set $f(G)$ to $f$  and considered the initial conditions $\mathcal{G}_{in}=10^{-5}$, $\mathbf{G}_{in}=10^{-5}$ and $\Delta_{in}=10^{-5}$ to find numerical solutions of Eq. \ref{eq51} through to Eq. \ref{eq56} which are  presented  in Fig. \ref{Fig4}  through to Fig. \ref{Fig5}. From the plots, the energy overdensity decay with increase in redshift.
 \clearpage
  The  Eq. \ref{eq51} through to Eq. \ref{eq56} are presented for the generalized Chaplygin gas model. The original Chaplygin gas  model can be obtained by setting $\alpha=1$  in eq. \ref{eq51} through to eq. \ref{eq56} and  its evolution equations are presented as
  \begin{eqnarray}
 &&\Delta'_{m}=\frac{1}{(1+z)H}\Big(\Big(1-\frac{1}{(1+w_{t})\rho_{t}}\Big)\frac{(1+z)HGG'}{\theta^{2}}\Big)Z\nonumber\\&&+\frac{1}{(1+z)H} \frac{1}{(1+w_{t})\rho_{t}}w_{chp}\rho_{chp}\Delta_{chp}+\frac{1}{(1+z)H}\frac{1}{(1+w_{t})\rho_{t}}\frac{Gf''}{\theta}\mathsf{G}\nonumber\\&&
-\frac{1}{(1+z)H}\frac{1}{(1+w_{t})\rho_{t}}\Big(\frac{1}{2}(1-f'-Gf'')G(1+z)HG'f'''-\frac{(1+z)HG'f''}{G^{2}}\Big)\mathcal{G},
 \label{eq57}\\
&& Z'=\frac{1}{(1+z)H}\Big(\frac{1}{2}\rho_{d}+\frac{1}{3}\theta^{2}\Big)\Delta_{m}\nonumber\\
&&-\frac{1}{(1+z)H}\Big[-\frac{1}{2}(1+3w_{chp})\rho_{chp}+\big(\frac{1}{3}\theta^{2}+\frac{1}{2}(1+3w_{t})\rho_{t}\big)\frac{w_{chp}\rho_{chp}}{(1+w_{t})\rho_{t}}\nonumber\\
&&2\Big[\frac{3}{2}\rho_{chp}-\big(\frac{1}{3}\theta^{2}+\frac{1}{2}(1+3w_{t})\rho_{t}\big)\frac{\rho_{chp}}{(1+w_{t})\rho_{t}}\Big]w_{chp}\Big]\Delta_{chp}\nonumber\\&&-\frac{1}{(1+z)H}\Big[1-3f''-12H^{3}b_{1}f^{iv}+12H\Big(H(b_{2}f'''+\frac{b_{3}}{b_{1}}f''+b_{1}^{2}f^{iv})+2H^{3}f''\Big)\nonumber\\
&&+\big(\frac{1}{3}\theta^{2}+\frac{1}{2}(1+3w_{t})\rho_{t}\big)\frac{1}{(1+w_{t})\rho_{t}}\Big(\frac{1}{2}(1-f'-Gf'-Gf'')+Gb_{1}f'''+\frac{b_{1}f''}{G^{2}}\Big)\Big]\mathcal{G}\nonumber\\
&&-\frac{1}{(1+z)H}\Big(24b_{1}f'''-12H^{3}f'''+\big(\frac{1}{3}\theta^{2}+\frac{1}{2}(1+3w_{t})\rho_{t}\big)\frac{Gf''}{\theta(1+w_{t})\rho_{t}}\Big)\mathbf{G}\nonumber\\
&&-\frac{1}{(1+z)H}\Big[8H\left(f''b_{2}+f'''b_{1}^{2}+3Hf'-\frac{3}{2}f'''Hb_{1}\right)-\frac{2}{3}\theta\nonumber\\&&-\big(\frac{1}{3}\theta^{2}+\frac{1}{2}(1+3w_{t})\rho_{t}\big)\frac{Gb_{1}}{\theta^{2}(1+w_{t})\rho_{t}}\Big]Z
\label{eq58}\\
 && \mathcal{G}'= -\frac{1}{(1+z)H}\Big(1-\frac{b_{1} G f''}{(1+w_{t})\rho_{t} \theta}\Big)\mathsf{G} -\frac{1}{(1+z)H}\frac{b_{1}^{2} G}{(1+w_{t})\rho_{t} \theta^{2}} Z  \nonumber\\&&-\frac{1}{(1+z)H}\frac{ b_{1} \rho_{ch}w_{ch}}{(1+w_{t})\rho_{t}}\Delta_{ch}
 +\Big(\frac{1}{2(1+z)H}\frac{b_{1}}{(1+w_{t})\rho_{t}}\Big(f'+Gf''-1\Big)\nonumber\\&&+\frac{1}{2(1+z)H}\frac{b_{1}^{2}}{(1+w_{t})\rho_{t}}\Big(Gf''+\frac{f''}{\theta^{2}}\Big)\Big)\mathcal{G}
 \label{eq59}\\
 && \mathsf{G}'=-\frac{1}{(1+z)H}\Big(\frac{b_{3}}{b_{1}}-\frac{b_{1}{G}f''}{(1+w_{t})\rho_{t}\theta}\Big)\mathsf{G}+\Big[\frac{1}{(1+z)H}\frac{b_{2}b_{1}{G}}{(1+w_{t})\rho_{t}\theta^{2}}\Big]Z-\Big[\frac{1}{(1+z)H}\frac{ b_{2} \rho_{ch}w_{ch}}{(1+w_{t})\rho_{t}}\Big]\Delta_{ch}\nonumber\\&&-\frac{1}{(1+z)H}\Big[\frac{b_{2}}{2(1+w_{t})\rho_{t}}(f'+Gf''-1)-\frac{b_{2}b_{1}}{(1+w_{t})\rho_{t}}(Gf''+\frac{f''}{\theta^{2}})\Big]\mathcal{G}
 \label{eq60}
 \end{eqnarray}
\begin{eqnarray}
&&\Delta'_{G}=-\frac{1}{(1+z)H}\Big[-\frac{4H\theta}{\rho_{G}}\Big[2H b_{1}f''' +(2c_{1}-2H^{2})f''\Big]\nonumber\\&&+\theta \Big(4H^{2}f''b_{2}+4H(2c_{1}-H^{2})f''b_{1}\Big)\Big(\frac{Gf''}{\theta(1+w_{t})\rho_{t}\rho_{G}}\Big)\Big]\mathsf{G}\nonumber\\&&
-\frac{1}{(1+z)H}\Big[4H\Big[H(b_{2}f'''+\frac{f''b_{2}}{b_{2}}+b_{2}^{2}f''')+(2c_{1}-H^{2})f''\Big]\nonumber\\&&+\theta\Big(4H^{2}(f''b_{2}+f'''b^{2}_{1}))+4H(2c_{1}-H^{2})f''b_{1}\Big(\frac{\frac{1}{2}+Gb_{1}f''+\frac{b_{1}f'''}{\theta^{2}}}{(1+w_{t})\rho_{t}\rho_{G}}\Big)\Big]\mathcal{G} +\nonumber\\
&&-\frac{1}{3((1+z)H)\rho_{G}}\Big[\Big(2H(f''b_{2}+f'''b_{1}^{2})-8Hf''b_{1}(\frac{G}{12H^{3}}+3H)\nonumber\\&&+(2c_{1}-H^{2})f''b_{1}-3\Big(4H^{2}(f''b_{2}+f'''b^{2}_{1})+4H(2c_{1}-H^{2})f''b_{1}\Big)\nonumber\\
&&+\theta(4H^{2}(f''b_{2}+f'''b^{2}_{1})-4H(2c_{1}-H^{2})f''b_{1})(\frac{Gb_{1}\theta^{2}}{(1+w_{t})\rho_{t}\rho_{G}})\Big)\nonumber\\ &&+3\theta(4H^{2}(f''b_{2}+f'''b^{2}_{1})+4H(2c_{1}-H^{2})f''b_{1})\Big(\frac{(1-f'-Gf'')Gb_{1}}{\theta^{2}(1+w_{t})\rho_{t}}\Big)\Big]Z
\nonumber\\
&&-\frac{3}{(1+z)\rho_{G}}\Big(4H^{2}(f''b_{2}+f'''b^{2}_{1})+4H(2c_{1}-H^{2})f''b_{1}\Big) \Delta_{G}\nonumber\\&&+\frac{6}{(1+z)} \Big(4H^{2}(f''b_{2}+f'''b^{2}_{1})+4H(2c_{1}-H^{2})f''b_{1}\Big)\Big(\frac{w_{ch}\rho_{ch}}{(1+w_{t})\rho_{t}\rho_{G}}\Big)w_{ch}\Delta_{ch}
\label{eq61}\\
 &&\Delta^{'}_{ch}= \frac{1}{(1+z)H}\Big[1+\frac{Gb_{1}}{\theta(1+w_{t})\rho_{t}}\Big](1+w_{ch})Z-\frac{3}{(1+z)}\Big[(1+\rho_{ch})\frac{ (1+w_{ch})}{(1+w_{t})\rho_{t}}w_{ch}\rho_{ch}-2w_{ch}\Big]\Delta_{ch}\nonumber\\&&-\frac{\theta (1+w_{ch})}{(1+z)(1+w_{t})H\rho_{t}}\Big[\frac{1}{2}(1-f'-Gf'')+Gb_{1}f''+\frac{b_{1}f'''}{\theta^{2}}\Big]\mathcal{G}-\frac{\theta (1+w_{ch})Gf''}{(1+z)(1+w_{t})H\rho_{t}}\mathsf{G}
 \label{eq62}
\end{eqnarray}
which admit the numerical solutions presented in fig. \ref{Fig6}.
\begin{figure}
  \includegraphics[width=140mm,height=95mm]{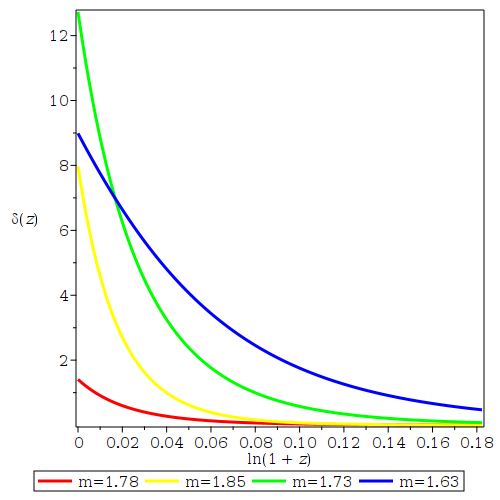}
  \caption{Plot of energy density perturbations versus redshift of Eq. \ref{eq57}-\ref{eq62} for different values of $m$ in the dust dominated Universe for long wavelength mode $k=0.00001$ in a generalized Chaplygin model $\alpha=1$.}
  \label{Fig6}
 \end{figure}
  We considered the initial conditions $\mathcal{G}_{in}=10^{-5}$, $\mathbf{G}_{in}=10^{-5}$ and $\Delta_{in}=10^{-5}$ to find numerical solutions of Eq. \ref{eq57} through to Eq. \ref{eq62}.  The energy overdensity fluctuations decay with increase in redshift.
 \section{Discussion and Conclusions}\label{section5}
The current treatment is focused on a mixture of matter fluid, chaplygin gas and Gauss-Bonnet fluid as non-interacting entities to study cosmological perturbations in $f(G)$ gravity using $1+3$ covariant Gauge invariant formalism. After presenting the covariant form of field equations and the kinematic quantities in the presence of Gauss-Bonnet invariant, we define matter fluids, chaplygin gas fluids and Gauss-Bonnet fluids to derive first order evolution equations for the defined gradient variables. Since our main interest lies in the structure formation of the Universe, we use scalar decomposition method to decompose the evolotion equations into their scalar parts believed to be responsible for the spherical clustering of large scale structure. These  scalar perturbation equations are considered as input to study the energy overdensity fluctuations in a dust-Chaplygin gas-Gauss-Bonnet system by applying harmonic decomposition method. After we transform the linear perturbation equations into redshift space for further analysis. We apply both the generalized and original chaplygin gas together with the  $f(G)$ model to find numerical results and analyse how the energy density fluctuations evolve with  redshift. We considered the short wavelength limits of the perturbation equations in the Universe dominated by a mixture of dust-chaplygin gas-Gauss-Bonnet fluids system and we depict that the energy overdensity perturbations decay with increase in redshift. From all plots, we observe that the decay of matter density contrast is fast decaying with redshift compared to the matter-Chaplygin-Gauss-Bonnet fluids system, since the universe  is believed to become less structured in the accelerated phase of the universe evolution. We have considered different initial conditions such as $\mathcal{G}_{in}=10^{-5}$, $\mathbf{G}_{in}=10^{-5}$ and $\Delta_{in}=10^{-5}$ to find numerical solutions of different perturbation equations.  The use of initial conditions constrained the amplitude of matter energy densities but does not alter the behaviour of the curves, but the values of the parameter $m$, $k$ and other considered parameters changes the behavior of the curves. From all the plots, there is no oscilatory behaviours observed.\\
\hfill\\
 \textbf{Some of the specific highlights of the present work include:}
 \begin{itemize}
  \item We have derived the linear perturbation equations eq. \ref{eq29}-\ref{eq34} in the context of $f(G)$ gravity, believed to contribute to the large scale structure formation.
  \item The matter energy density fluctuations couples with the energy density fluctuations of the chaplygin gas and that of Gauss-Bonnet fluids which then decouple for the GR limits.
  \item In the analysis stage, we considered short wavelength and long wavelength modes for the energy overdensity fluctuations in a Universe dominated by a mixture of a dust-chaplygin gas-Gauss-Bonnet fluids by combining chaplygin gas and the $f(G)$ models which can mimics the $\Lambda$CDM  for the $f(G)=G$ case.
  \item We solved the perturbation equations using numerical method and the numerical results of the energy overdensity fluctuations are presented in Fig. \ref{Fig1} GR limits and Fig. \ref{Fig2}-\ref{Fig6} for the combination of chaplygin gas and $f(G)$ model in the short and long wavelength modes. From these figures the energy overdensity decay with redshift for the considered range of parameter $m$.
  \item From the plots, the numerical results of the perturbation equations are sensible to the parameter $m$. As $m$ changes, we notice an change in the amplitudes of the energy overdensity fluctuations whih reduces to the GR limits for $m=1$ but the results do not depend much on the parameter $\alpha$.
  \item Under the consideration of both original and generalized Chaplygin gas models, it was observed that there is no significant difference in the behavior of the results.
 \end{itemize}In general, the combination of chaplygin gas models and $f(G)$ model offers an alternative for large scale structure formation scenarios even though there is a variation in the growth of amplitudes of the energy overdensity fluctuations which can be beneficial for constraining the parameters of the models by referring to the  observational data which provide different cosmological scenarios that are consistent with the $\Lambda$CDM limits. We look forward to undertake this aspect of the task to  different chaplygin gas models as well as $f(G)$ models in the $1+3$ covariant cosmological perturbations framework for the future work.\\
\section*{Appendix}
\section*{Useful Linearised Differential Identities}
For all scalars $f$, vectors $V_a$ and tensors that vanish in the background,
$S_{ab}=S_{\la ab\ra}$, the following linearised identities hold:
\begin{eqnarray}
\left(\D_{\la a}\D_{b\ra}f\right)^{.}&=&\D_{\la a}\D_{b\ra}\dot{f}-\sfrac{2}{3}\Theta\D_{\la a}\D_{b\ra}f+\dot{f}\D_{\la a}A_{b\ra}\label{a0}\;,\\
\ep^{abc}\D_b \D_cf &=& 0 \label{a1}\;, \\
\ep_{cda}\D^{c}\D_{\la b}\D^{d\ra}f&=&\ep_{cda}\D^{c}\D_{( b}\D^{d)}f=\ep_{cda}\D^{c}\D_{ b}\D^{d}f=0\label{a2}\;,\\
\D^2\left(\D_af\right) &=&\D_a\left(\D^2f\right)
+\sfrac{1}{3}\tl{R}\D_a f \label{a4}\;,\\
\left(\D_af\right)^{\rd} &=& \D_a\dot{f}-\sfrac{1}{3}\Theta\D_af+\dot{f}A_a
\label{a14}\;,\\
\left(\D_aS_{b\cdots}\right)^{\rd} &=& \D_a\dot{S}_{b\cdots}
-\sfrac{1}{3}\Theta\D_aS_{b\cdots}
\label{a15}\;,\\
\left(\D^2 f\right)^{\rd} &=& \D^2\dot{f}-\sfrac{2}{3}\Theta\D^2 f
+\dot{f}\D^a A_a \label{a21}\;,\\
\D_{[a}\D_{b]}V_c &=&
-\sfrac{1}{6}\tl{R}V_{[a}h_{b]c} \label{a16}\;,\\
\D_{[a}\D_{b]}S^{cd} &=& -\sfrac{1}{3}\tl{R}S_{[a}{}^{(c}h_{b]}{}^{d)} \label{a17}\;,\\
\D^a\left(\ep_{abc}\D^bV^c\right) &=& 0 \label{a20}\;,\\
\label{divcurl}\D_b\left(\ep^{cd\la a}\D_c S^{b\ra}_d\right) &=& {\ts{1\over2}}\ep^{abc}\D_b \left(\D_d S^d_c\right)\;,\\
\text{curlcurl} V_{a}&=&\D_{a}\left(\D^{b}V_{b}\right)-\D^{2}V_{a}+\sfrac{1}{3}\tl{R}V_{a}\label{curlcurla}\;,
\label{a21}
\end{eqnarray}
\section*{Used parameters in redshif space}
\begin{eqnarray}
&&H=\frac{2m}{3}(1+z)^{\frac{3}{2m}} \\
  &&\dot{H}=c_{1}=-\frac{2m}{3}(1+z)^{\frac{3}{m}}\\
 &&\dot{G}=b_{1}=\frac{256}{9}m^{3}\Big(1-\frac{2m}{3}\Big)(1+z)^{\frac{15}{2m}} \\
 && \ddot{G}=b_{2}=1280m^{3}\Big(\frac{2m}{3}-1\Big)(1+z)^{\frac{9}{m}}\\
 && \dddot{G}=b_{3}=\frac{2560}{3}m^{3}\Big(1-\frac{2m}{3}\Big)(1+z)^{\frac{21}{2m}}\\
  &&\dot{f}=-(1+z)Hf'.
  \end{eqnarray}
\section*{Acknowledgments}
 We thank the anonymous reviewer(s) for the constructive comments towards the significant improvement of this manuscript. The authors acknowledge the financial support from UR-SIDA Grant No. DVC-AAR506/2022\;. AM acknowledges the  support from International Science Program (ISP) to Rwanda Astrophysics, Space and Climate Science Research Group (RASCSRG), University of Rwanda grant No: RWA01 and to East African Astrophysics Research Network(EAARN) grant No: Afro:05. AM also acknowledges the hospitality of the department of Physics of the University of Rwanda, where  part of this work was conceptualise and completed.
 \small
\bibliographystyle{unsrt}

 \noindent
{\color{blue} \rule{\linewidth}{1mm} }
\end{document}